\documentstyle[prl,aps,epsf,twocolumn]{revtex}

\draft
\begin{document}

\wideabs{
\title{Sympathetic cooling of bosonic and fermionic Lithium gases towards quantum
degeneracy}
\author{F.\,Schreck , G.\,Ferrari, K. L.\,Corwin, J.\,Cubizolles,
L.\,Khaykovich, M.-O.\,Mewes and C.\,Salomon}
\address{Laboratoire Kastler Brossel, Ecole Normale Sup\'erieure, 24 rue Lhomond, 75231 Paris CEDEX 05, France}
\date{\today}
\maketitle

\begin{abstract}
Sympathetic cooling of two atomic isotopes is experimentally investigated. Using forced
evaporation of a bosonic $^7$Li gas in a magnetic trap, a sample of $3\,10^5$ $^6$Li
fermions has been sympathetically cooled to $9(3)\, \mu$K, corresponding to $2.2(0.8)$
times the Fermi temperature. The measured rate constant for $2$-body inelastic collisions
of $^7$Li $|2,2\rangle$ state at low magnetic field is $1.0^{+0.8}_{-0.5} \,
10^{-14}$\,cm$^3$s$^{-1}$.
\end{abstract}

 }

In atomic physics, the combination of laser cooling and evaporative cooling in magnetic
traps has been successfully used to reach Bose-Einstein condensation in dilute vapors
\cite{Anderson95,EnricoFermi99}. These techniques, however, are not universal. Producing
laser light at appropriate wavelengths is sometimes difficult and laser cooling of
molecules remains a challenge. Relying on elastic collisions \cite{Hess86}, evaporative
cooling fails for fermions at low temperature. Indeed no {\it s}-wave scattering is
allowed for identical fermions and when the temperature $T$ decreases, the {\it p}-wave
cross-section vanishes as $T^2$ \cite{Demarco99}.

Sympathetic cooling allows one to overcome these limitations. It uses a buffer gas to cool
another species via collisions and was first proposed for two-component plasmas
\cite{plasmas}. Often used for cooling ions confined in electromagnetic traps
\cite{Wineland80,Wineland86}, it has been applied recently to cool neutral atoms and
molecules via cryogenically cooled Helium \cite{Doyle97}. Sympathetic cooling using
$^{87}$Rb atoms in two different internal states has led to the production of two
overlapping condensates \cite{Myatt97}. For fermions, the {\it s}-wave scattering
limitation was overcome by using two distinct Zeeman substates, both of which were
evaporatively cooled. This method has been used to reach temperatures on the order of
$\sim 300\,$nK\,$ \sim 0.4
\,T_{\rm F}$ \cite{Jin99}, where $T_{\rm F}$ is the Fermi temperature  below which
quantum effects become prominent. Cold collisions between $^6$Li and $^7$Li were analyzed
theoretically in \cite{VanAbeleen97} and it was predicted that sympathetic cooling of
$^6$Li by contact with $^7$Li should work efficiently.

In this Letter, we report on the experimental demonstration of sympathetic cooling of
$^6$Li fermions via collisions with evaporatively cooled $^7$Li bosons in a magnetic trap.
Both species were cooled from $2$ mK to $\sim 9(3)\,\mu$K, corresponding to $T\sim
2.2(0.8)\, T_{\rm F}$ where $T_{\rm F}=(\hbar \bar{\omega}/k_B)(6 N)^{1/3}$,
$\bar{\omega}$ is the geometric mean of the three oscillation frequencies in the trap and
$N$ is the number of fermions. For these experimental conditions, $T_{\rm F} \sim
4\,\mu$K. This method represents a crucial step towards the production of a strongly
degenerate Fermi gas of $^6$Li and the study of its optical and collisional properties
\cite{Gamma,Ferrari99}. It also opens the way to interesting studies on mixtures of Bose
condensates and Fermi gases \cite{Smith99,Molmer99,Geist99}. Finally $^6$Li is considered
a good candidate for the observation of BCS transition \cite{Stoof96,Combescot99,Stoof99}.

\begin{figure}[t]
\begin{center}
\epsfxsize=8cm
\leavevmode
\epsfbox{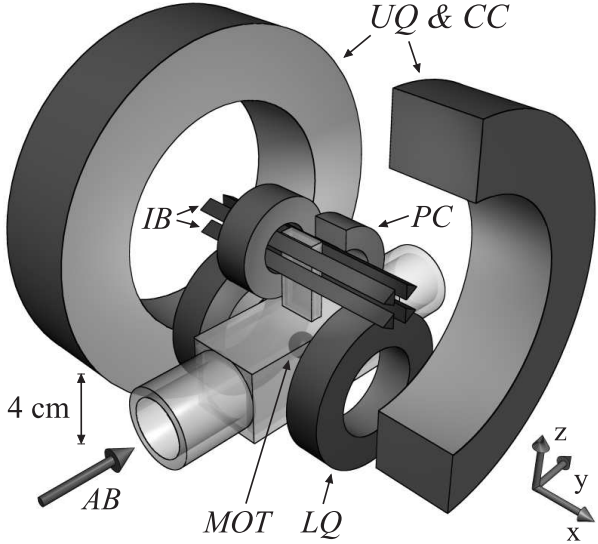}
\caption{
\label{fig:setup}
Experimental set-up. Both lithium isotopes are collected from a slow atomic beam ({\it
AB}) in a magneto-optical trap ({\it MOT}) at the center of a glass cell. Atoms are
magnetically elevated using lower quadrupole ({\it LQ}) and upper quadrupole ({\it UQ})
coils into a small appendage. At this site, a strongly confining Ioffe-Pritchard trap
consisting of 4 Ioffe bars ({\it IB}), two pinch coils ({\it PC}) and two compensation
coils ({\it CC}) allows evaporative cooling of $^7$Li to quantum degeneracy and
sympathetic cooling of $^6$Li-$^7$Li mixtures. }
\end{center}
\end{figure}

A sketch of our apparatus is shown in Fig.\ref{fig:setup}. First the $^7$Li and $^6$Li
isotopes are simultaneously captured from a slowed atomic beam and cooled in a
magneto-optical trap ({\it MOT}) at the center of a Vycor glass cell \cite{TIMOT}. On top
of this cell is located a small appendage of external dimensions $20\times 7 \times
40$\,mm and a wall thickness of $2$\,mm. The small dimension along $y$ permits the
construction of a strongly confining Ioffe-Pritchard ({\it IP}) trap using
electro-magnets. The 2D quadrupole field is created by four copper Ioffe bars ({\it IB}),
each consisting of three conductors running at a maximum current of $700$\,A. Axial
confinement is provided by two pinch coils ({\it PC}) in series with two compensation
coils ({\it CC}) at $500$\,A in order to reduce the bias magnetic field. For these
currents, the radial gradient is $2.38$\, kG/cm and the axial curvature is
$695$\,G/cm$^2$. With a bias field of $15$\,G, the trap frequencies for $^7$Li are
$\omega_{\rm rad}/2\pi= 2.57(4)$\,kHz and $\omega_{\rm ax}/2\pi=118(1)$\,Hz. The
background-limited trap lifetime is 130\,s. Atoms are transferred from the {\it MOT}
region to the {\it IP} trap in a magnetic elevator quadrupole trap. The elevator consists
of lower quadrupole ({\it LQ}) coils, (identical with the {\it MOT} coils) and upper
quadrupole ({\it UQ}) coils centered on the {\it IP} trap axis, 50\,mm above the LQ trap
center. The current ratio between the two sets of coils determines the center of the
resulting quadrupole trap, allowing the atoms to be lifted by adjusting this ratio.

This trap design allows both large compression and high atom numbers, two crucial
parameters for evaporative cooling. This is particularly important for lithium atoms since
the {\it s}-wave cross-sections at zero energy are at least 16 times smaller than for
$^{87}$Rb. The scattering length for $^7$Li in state $|F=2, m_F=2\rangle$ used in our
experiments is $-27 a_0$ \cite{Abraham97}. Finally, all coil currents can be switched off
rapidly, allowing the potentials to be turned off non-adiabatically for time-of-flight
absorption imaging. In contrast, early BEC experiments with $^7$Li used a permanent magnet
trap \cite{Hulet95}.
We describe now the main steps of our sympathetic cooling experiments (parameters are
summarized in Table 1). $6\,10^{9}$ $^7$Li and $1.6\,10^{8}$ $^6$Li atoms are captured in
$\sim 1$ minute in the {\it MOT}. The relative numbers of $^7$Li and $^6$Li can be
adjusted by changing the light intensity tuned to each isotope with the laser set-up
described in \cite{FerrariMOPA}. Both isotopes are cooled to $\sim 0.8$\,mK, optically
pumped to the upper hyperfine state, F=2 (resp. F=3/2) and captured in the {\it LQ} trap
with an axial gradient of 400\,G/cm. Characterization of the LQ trap with $^7$Li revealed
two time scales for trap losses: a fast one (100\,ms) that we attribute to spin relaxation
and a slow one (50 s) due to Majorana transitions near the trap center. Just before the
magnetic elevator stage about $40\%$ of both isotopes remain trapped.

The transfer to the {\it UQ} trap is done by increasing the current in the {\it UQ} coils
to 480 A in 50 ms and ramping off the current in the {\it LQ} coils in the next 50 ms.
Nearly mode-matched transfer into the {\it IP} trap is accomplished by simultaneously
switching on the {\it IB} and the pinch coils {\it PC} while switching off the {\it UQ}
coils. The transfer efficiency from {\it LQ} trap to the {\it IP} trap is $15\%$ and is
limited by the energy cut due to the narrow dimension of the appendage (3\,mm) in radial
direction. High energy atoms hit the glass cell and are lost. A wider appendage would
allow a higher transfer efficiency but at the expense of a reduced radial gradient. Our
simulation shows that the choosen size is optimum with respect to the initial collision
rate in the {\it IP} trap for a temperature of $1\,$mK in the {\it LQ} trap.

After compressing the {\it IP } trap to maximum currents and reducing the bias field to
$15$ G, we obtain $2.5\,10^{8}$ $^7$Li atoms and $1.8\,10^{7}$ $^6$Li atoms at a
temperature of $ 7(3)$ mK. In these conditions, we have been unsuccessful in reaching
runaway evaporation (i.e increase of collision rate) on $^7$Li in presence or in absence
of $^6$Li. Because the $|2,2\rangle$ state of $^7$Li has a negative scattering length, as
the collision energy increases, the scattering cross-section falls to zero. This occurs at
$T_o=4\, $mK, i.e. within the {\it s}-wave energy range \cite{Dalibard99}. To overcome
this limitation, we apply a stage of 1-dimensional Doppler cooling of $^7$Li in the {\it
IP } trap with a bias field of 430 G using a $\sigma^+-\sigma^+$ laser standing wave
aligned along the $x$-axis. Each beam has an intensity of $25\,\mu$W/cm$^2$ and is detuned
about one natural linewidth below the resonance in the trap. In 1\,s of cooling, the
temperature in all three dimensions drops by a factor of $4$ and the loss of atoms is
$15\%$. In the compressed {\it IP } trap, the temperature is now $\sim 2\, $mK,
sufficiently below $T_0$. Thus at the start of the evaporation, the collision rate is
$\sim 15\,$ s$^{-1}$, i.e $2000$ times the background gas collision rate.

Absorption images are taken with $10\,\mu$s exposure time, immediately after the trap is
turned off and before the cloud can expand. Independent laser systems provide isotope
selective probe beams. From these images, the total number of atoms is found, to an
accuracy of a factor of 2. The temperature is then deduced from a fit of the cloud size in
the axial direction by a Gaussian of standard deviation $\sigma_{\rm ax}$ and the measured
oscillation frequency $\omega_{\rm ax}$. This temperature measurement agrees to within
15\,\% with a time-of-flight measurement of the kinetic energy.

\begin{figure}[t]
\begin{center}
\epsfxsize=8cm
\leavevmode
\epsfbox{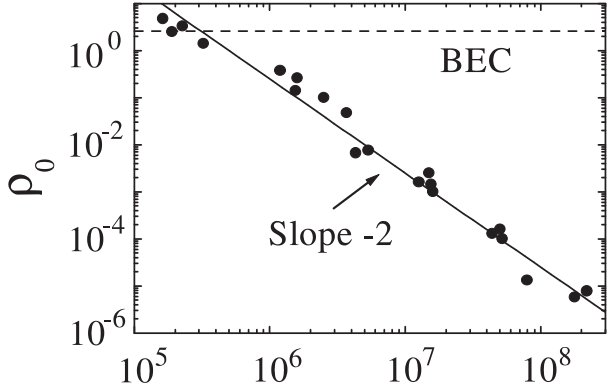}
\caption{$^7$Li peak phase-space density {\it vs} number of atoms $N$ during single-species evaporation.
\label{fig:li7}}
\end{center}
\end{figure}

\begin{figure}[t]
\begin{center}
\epsfxsize=8cm
\leavevmode
\epsfbox{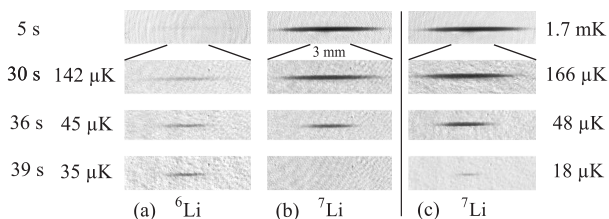}
\caption{Images of $^6$Li and $^7$Li atom clouds at various stages of
  sympathetic cooling (a, b) and
during single-species evaporation (c) with identical initial numbers of $^7$Li atoms. Top
images are 1 cm long, the others 3 mm. Temperatures of $^6$Li (resp. $^7$Li alone) are
given on the left (right).
\label{fig:li6}}
\end{center}
\end{figure}

Evaporation is performed exclusively on $^7$Li. We apply a tunable microwave field near
the $^7$Li hyperfine transition at $803.5\,$MHz to couple the $|F=2, m_F=2\rangle $
trapped state to the $|F=1, m_F=1\rangle $ untrapped state. To verify the effectiveness of
the evaporation process, we first cool $^7$Li alone in the trap down to the regime of
quantum degeneracy; after $15$\,s of compression of the cloud with a microwave knife fixed
at $5.4$\,mK, we have $2\,10^8$ atoms at a temperature of $1.2$\,mK \cite{note}. We then
lower the microwave cut energy from this value to $\sim 5\,\mu$K in
 45\,s. The central phase space density is $\rho_{0} =n_{0}
\Lambda^3 $ where $\Lambda= h(2\pi m k_B T)^{-1/2}$ and $n_{0}$ is the peak density. In Fig.
\ref{fig:li7}, $\rho_{0}$ is plotted {\it vs} atom number at various stages of the
evaporation scan. $\rho_{0}(N)$ is well fitted by a line of slope $-2$ on a log-log scale,
indicating a nearly constant collision rate during forced evaporation. We reached
phase-space densities exceeding $2.6$ with $5\,10^4$ atoms at $1.2\,
\mu$K, meeting the condition for Bose-Einstein condensation. However, due to the effective
attractive interaction, the trapped condensate is expected to be stable
 only for atom numbers up to a critical value \cite{Ruprecht95,Bradley97} which, in our
trap geometry, is $\sim 600$. Our current imaging system is unable to detect such a small
sample. We also observe that the trap lifetime is reduced to $22(4)\,$s at a peak density
of $4\, 10^{12}\,$at/cm$^3$ and a temperature of $21\,\mu$K, well above BEC. We attribute
this reduction to dipolar relaxation and obtain a dipolar relaxation rate $\beta_{7-7}=
1.0^{+0.8}_{-0.5}
\, 10^{-14}$\,cm$^3$s$^{-1}$ in agreement with the theoretical value
$1.6\,10^{-14}$\,cm$^3$s$^{-1}$ predicted in \cite{Moerdijk96} and comparable to that
measured in high magnetic field \cite{Gerton99,beta}.
Using a similar evaporation ramp lasting 40 s with $2.5\,10^6$ $^6$Li atoms and
$3.2\,10^8$ $^7$Li atoms, sympathetic cooling of fermionic $^6$Li is clearly shown in
Fig.\ref{fig:li6}. The displayed images recorded at various stages of the evaporation ramp
for mixtures (a and b) or for $^7$Li alone with identical initial number (c). In (a), the
optical density of the $^6$Li cloud is seen to increase considerably because of the
reduction in size without apparent loss of atoms, a signature of sympathetic cooling.
Comparisons of the cloud sizes between (a) and (b) indicate that $^6$Li and $^7$Li are in
thermal equilibrium, except for the end of the evaporation. At 39 s, $^6$Li is at
$40\,\mu$K, $^7$Li is not detectable (b), while $^7$Li alone is at $20\,\mu$K (c). This
indicates that between 36 and 39 s, the numbers of $^6$Li and $^7$Li have become equal.
Beyond this point, the temperature of $^6$Li no longer decreases significantly. The
thermal capacity of $^6$Li soon exceeds that of $^7$Li, resulting in heating and loss of
$^7$Li during the final stage of forced evaporation (b). Under the same conditions but
without $^6$Li, normal evaporative cooling of $^7$Li proceeds to a temperature of
$18\,\mu$K (c) which is below that obtained for $^6$Li at 39 s, $35\,\mu$K (a).

\begin{figure}[t]
\begin{center}
\epsfxsize=8cm
\leavevmode
\epsfbox{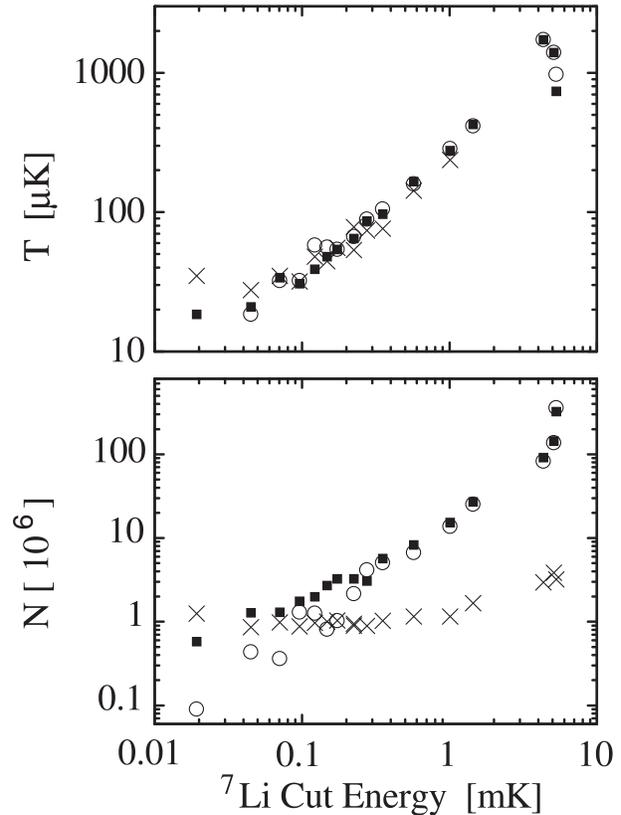}
\caption{$^6$Li (crosses) and $^7$Li (open circles) temperatures and numbers as a function
of the $^7$Li cut energy. During sympathetic cooling, temperatures are the same down to a
decoupling region below which $^6$Li is no longer cooled. Black squares show $^7$Li alone
with identical parameters. Uncertainty in the $x$-axis is about 0.01 mK.
\label{fig:charact}}
\end{center}
\end{figure}

More quantitatively, temperatures and numbers of $^6$Li and $^7$Li atoms as a function of
the microwave cut energy are plotted in Fig.\ref{fig:charact}. Above $40\,\mu$K, the
temperature of all three clouds are nearly identical, indicating that the collision
cross-section between the $^7$Li $|2,2\rangle$ and $^6$Li
$|\frac{3}{2},\frac{3}{2}\rangle$ is not significantly smaller than that of the $^7$Li
$|2,2\rangle$ with itself. This is consistent with the prediction that the scattering
length between these states of $^6$Li and $^7$Li is (40.8 $\pm$ 0.2) $a_0$
\cite{VanAbeleen97}. In addition, close thermal contact implies that the microwave knife
acts the same on the $^7$Li cloud with or without $^6$Li present, as expected because the
$^6$Li number is initially a very small fraction of the $^7$Li number. The ratio between
the measured temperatures and the microwave cut energy is $4$. At a temperature around
$35\,\mu $K, the number of $^7$Li atoms ($N_7$) has been reduced to the number of $^6$Li
atoms ($N_6$). As a result the $^6$Li cloud is no longer cooled and remains at $35\,\mu$K.

For classical gases, the decoupling temperature $T_D$ is reached when $N_7 \approx N_6$.
In the case of $^7$Li alone, $T_7$ is approximately proportional to $N_7$ and so $T_D
\propto N_6$. Since $T_{\rm F} \propto N_6^{1/3}$, the degeneracy parameter
$T_D/T_{\rm F} \propto N_6^{2/3}$, and Fermi degeneracy can be approached by reducing the
number of $^6$Li. Doing this, the highest degeneracy reached after a complete sympathetic
cooling evaporation ramp was $T/T_{\rm F}=2.2(0.8)$ at a temperature of $9(3)\mu $K with
$1.3\,10^5$ $^6$Li atoms. At this stage we are limited by the detection efficiency of our
imaging system.

Finally an important question for studies of mixtures of these degenerate gases is the
possibility of inter-species loss mechanisms. We have searched for such losses by
recording the $^6$Li trap lifetimes in presence and in absence of $^7$Li atoms at a peak
density $n_0(^7$Li$)=2\,10^{11}$\,at/cm$^3$ and a common temperature of $530\,\mu$K. With
$n_0(^7$Li$)=4.6\,n_0(^6$Li$)$, the lifetimes were respectively $73(10)$\,s and
$73(8)$\,s, showing no significant difference. We deduce an upper limit for dipolar decay
rates, $\beta_{6-7}\leq 1\,10^{-13}$\,cm$^3$s$^{-1}$, $\beta_{6-6}\leq
4.6\,10^{-13}$\,cm$^3$s$^{-1}$.

In summary we have demonstrated sympathetic cooling of fermionic lithium via evaporation
performed on the bosonic Li isotope and obtained temperatures of $2.2(0.8)\, T_{\rm F}$.
Several improvements of the experiment should allow us to achieve full Fermi degeneracy
and investigate the properties of such a quantum gas and of degenerate Boson-Fermion
mixtures.

We are grateful for experimental assistance of Fabrice Gerbier and A. Sinatra, to ANDOR
technology for loaning us a camera and to J. Dalibard, C. Cohen-Tannoudji, G. Shlyapnikov
and D. Gu\'ery-Odelin for discussions. M.-O.\,M., F.\,S., and K.\,C. were supported by a
Marie-Curie Research fellowship of the EU, by a doctoral fellowship from the DAAD and by
MENRT. This work was partially supported by CNRS, Coll\`ege de France, DRED, and the EC
(TMR Network No. ERB FMRX-CT96-0002). Laboratoire Kastler Brossel is {\it Unit\'e de
recherche de l'Ecole Normale Sup\'erieure et de l'Universit\'e Pierre et Marie Curie,
associ\'ee au CNRS}.

\begin{table}
\begin{tabular}{l|c c|c }
 &\multicolumn{2}{c}{$^7$Li}&\multicolumn{1}{c}{$^6$Li}\\
 [0.5ex]                &N              &T\,$\-[$mK$\-]$    &N   \\
 [0.5ex]\hline
 Compressed MOT         &$6\,10^{9}$    &$0.8$    &$1.6\,10^{8}$  \\
 Lower Quadrupole       &$2.5\,10^{9}$  &$1$      &$8\,10^{8}$    \\
 Capture I-P trap       &$3.8\,10^{8}$  &$0.7$    &$1.3\,10^{7}$  \\
 Compressed I-P         &$3.2\,10^{8}$  &$2$      &$2.5\,10^{6}$  \\
 End symp. cool.        &$1.7\,10^{6}$  &$0.03$   &$1.2\,10^{6}$  \\
\end{tabular}
\caption{
\label{tab:TransferTable}
  Typical atom numbers and temperatures before and during sympathetic cooling}
\end{table}


\begin{thebibliography}{99}
\bibitem{note}
This compression phase was not used in $^6$Li cooling.
\bibitem{beta} the definition of $\beta$ is given by $\dot N= -\beta n_0 N/\sqrt{2}-\alpha N$
\end{thebibliography}
\end{document}